\documentclass[%
reprint,
superscriptaddress,
showpacs,
preprintnumbers,
 bibnotes,
 amsmath,amssymb,
 aps,
prb,
]{revtex4-1}

\usepackage{graphicx}
\usepackage{dcolumn}
\usepackage{bm}
\usepackage{color}

\begin{document}


\title{Soft vibrational mode associated with incommensurate orbital order in multiferroic CaMn$_7$O$_{12}$
}

\author{Xinyu~Du}
\affiliation{International Center for Quantum Materials, School of Physics, Peking University, Beijing 100871, China}
\author{Renliang~Yuan}
\affiliation{International Center for Quantum Materials, School of Physics, Peking University, Beijing 100871, China}
\author{Lian~Duan}
\affiliation{International Center for Quantum Materials, School of Physics, Peking University, Beijing 100871, China}
\author{Chong~Wang}
\affiliation{International Center for Quantum Materials, School of Physics, Peking University, Beijing 100871, China}
\author{Yuwen~Hu}
\affiliation{International Center for Quantum Materials, School of Physics, Peking University, Beijing 100871, China}
\author{Yuan~Li}
\email[]{yuan.li@pku.edu.cn}
\affiliation{International Center for Quantum Materials, School of Physics, Peking University, Beijing 100871, China}
\affiliation{Collaborative Innovation Center of Quantum Matter, Beijing 100871, China}

\begin{abstract}
We report inelastic light scattering measurements of lattice dynamics related to the incommensurate orbital order in $\mathrm{CaMn_7O_{12}}$. Below the ordering temperature $T_\mathrm{o} \approx 250 \,\mathrm{K}$, we observe extra phonon peaks as a result of Brillouin-zone folding, as well as a soft vibrational mode with a power-law $T$-dependent energy, $\Omega = \Omega_{0}(1 - T/T_{\mathrm{o}})^{1/2}$. This temperature dependence demonstrates the second-order nature of the transition at $T_\mathrm{o}$, and it indicates that the soft mode can be regarded as the amplitude excitation of the composite order parameter. Our result strongly suggests that the lattice degrees of freedom are actively involved in the orbital-ordering mechanism.
\end{abstract}


\maketitle


\section{\label{sec:intro}Introduction}

It is widely recognized that the orbital degrees of freedom are related to many fascinating physical phenomena in correlated-electron materials.\cite{1tokura2000orbital,2hotta2006orbital} The manifestation of this is particularly rich in perovskite-type manganites.\cite{1tokura2000orbital,3imada1998metal,4tokura2006critical,5maekawa2004physics} Of each Mn$^{3+}$ ion situated at the center of an oxygen octahedron, the four 3$d$ electrons tend to occupy three $t_{2g}$ and one $e_g$ orbital, forming a high-spin configuration that minimizes Hund's coupling energy. The $e_g$ orbital extending along direction(s) in which the negatively-charged ligand oxygen atoms are farthest away will have the lowest energy, hence a distortion of the oxygen octahedron lifts the $e_g$-orbital degeneracy and is energetically favorable. This is commonly known as the Jahn-Teller effect. Together with the orbital dependence of magnetic exchange interactions between neighboring Mn sites,\cite{5maekawa2004physics} it gives orbital physics a central role in the interplay among the lattice and electronic (charge, orbital, and spin) degrees of freedom.

Many of the orbital ordering phenomena in the manganites can be understood in a real-space picture,\cite{3imada1998metal,4tokura2006critical} in which the propagation of orbital state from one site to its neighbors is determined by cooperative Jahn-Teller effects \cite{6millis1996cooperative} and/or exchange interactions.\cite{7feiner1999electronic} It is thus no surprise that most orbital order in the manganites exhibits some form of ``lock-in'' with the crystal lattice,\cite{8murakami1998resonant,9murakami1998direct,10yamada1996polaron,11endoh1999transition,12radaelli1997charge} or with concomitant charge-ordering patterns.\cite{LarochellePRB2005,13beale2005orbital,14beale2010xray,15ulbrich2011evidence} In the latter case the Mn ions segregate themselves into 3+ and 4+ valence states, and because orbital order pertains only to the Mn$^{3+}$ sub-lattice, the ordering pattern develops on top of the charge order, usually in a commensurate fashion, even though the charge order itself may not be commensurate with the undistorted crystal lattice.\cite{LarochellePRB2005,13beale2005orbital,14beale2010xray,15ulbrich2011evidence}

\begin{figure}
\includegraphics[width=3.375in]{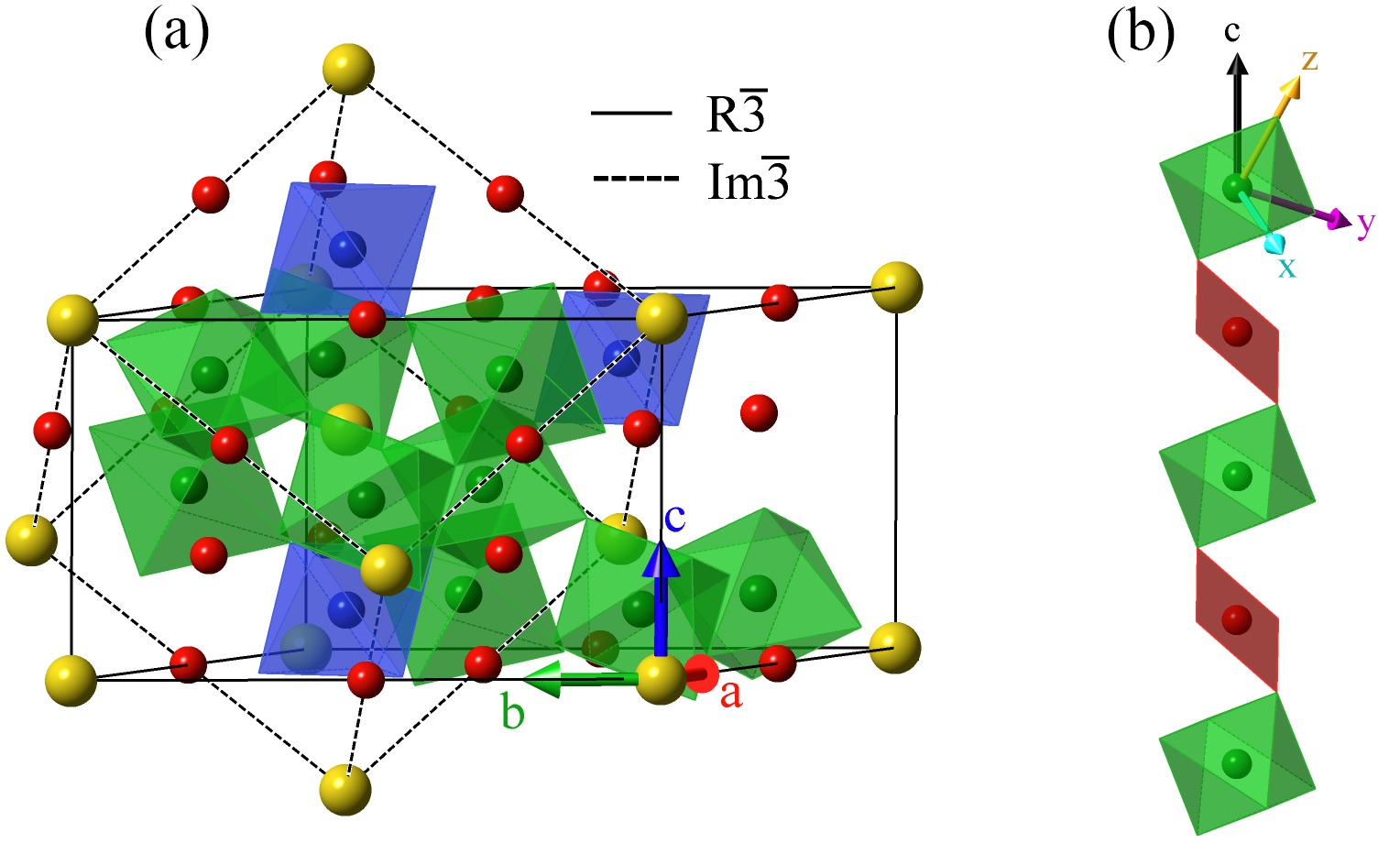}
\caption{\label{fig:structure}
(a) The crystal structure of $\mathrm{CaMn_7O_{12}}$ with hexagonal (solid lines) and pseudo-cubic (dashed lines) unit cells. Ca and Mn1 ions are shown in yellow and red spheres, respectively. Mn2 (green) and Mn3 (blue) ions are shown within oxygen octahedra. Oxygen atoms are not shown. (b): A Mn1-Mn2 chain along the hexagonal $c$-axis. Arrows indicate the local $xyz$ coordinate system for the definition of orbital states (see text).
}
\end{figure}

Recently, Perks \textit{et al.}\cite{16perks2012magneto} reported an unusual form of incommensurate orbital order in $\mathrm{CaMn_7O_{12}}$. Through a careful analysis of single-crystal X-ray diffraction data that builds on earlier reports of incommensurate lattice distortions below $\sim$ 250~K,\cite{17Przenioslo2004charge,18Slawinski2009modulation} the authors identified a continuous rotation of Mn$^{3+}$ $e_g$ orbital occupation between the local $3x^2 - r^2$ and $3y^2 - r^2$ states (Fig.~\ref{fig:structure}) propagating along the crystallographic $c$-axis. $\mathrm{CaMn_7O_{12}}$ exhibits multiferroic properties below $T_{\mathrm{N1}}=90$~K. An unprecedentedly large magnetically induced ferroelectric polarization \cite{19zhang2011multiferroic,20johnson2012giant,21lu2012giant} arises from a helical magnetic structure due to exchange striction,\cite{21lu2012giant} and it is suggested that the orbital order at higher temperature sets the stage for the helical magnetic structure to be stabilized.\cite{16perks2012magneto} These novelties motivated us to study the mechanism for the orbital ordering in $\mathrm{CaMn_7O_{12}}$ using Raman spectroscopy, which is sensitive to symmetry breaking via the observation of collective excitations. The key finding of our work is a soft vibrational mode which can be regarded as amplitude excitation of the composite orbital order parameter below $T_{\mathrm{o}}$. Our result constitutes the first direct observation of amplitude excitations associated with an orbital order, and it suggests the importance of lattice degrees of freedom in the formation of the orbital order in $\mathrm{CaMn_7O_{12}}$.

At high temperatures, $\mathrm{CaMn_7O_{12}}$ possesses the $\mathrm{AC_3B_4O_{12}}$ cubic structure (space group Im$\bar{3}$) which derives from simple perovskite $\mathrm{ABO_3}$ (Ref.~\onlinecite{22va}). The B-site Mn$^{3.25+}$ ions are situated at centers of corner-sharing distorted MnO$_6$ octahedra,\cite{23Bochu1980bond} whereas the C-site Mn$^{3+}$ (denoted as Mn1) ions are in MnO$_4$ rhombus configuration. Upon cooling through a first-order phase transition at $T_\mathrm{s} \approx 440$~K, one of the body diagonals of the cubic cell shrinks, making the crystal structure rhombohedral (space group R$\bar{3}$) at room temperature.\cite{23Bochu1980bond,24przenioslo2002phase} The rhombohedral $c$-axis in the hexagonal basis is parallel to the shortened cubic body diagonal (Fig.~\ref{fig:structure}a). Another aspect of this phase transition is that the four Mn$^{3.25+}$ ions in each formula unit undergo a charge order into three Mn$^{3+}$ (Mn2) and one Mn$^{4+}$ (Mn3) ions,\cite{23Bochu1980bond,24przenioslo2002phase} which occupy inequivalent sites in the rhombohedral structure. The MnO$_4$ rhombuses of Mn1 and the MnO$_6$ octahedra of Mn2 stack along the hexagonal $c$-axis in a corner-sharing fashion (Fig.~\ref{fig:structure}b).

The orbital order discovered by Perks \textit{et al.} features orbital occupations on Mn2 sites that continuously rotate between $3x^2 - r^2$ and $3y^2 - r^2$ states in the local coordinate system (Fig.~\ref{fig:structure}b) as one moves along the hexagonal $c$-axis, with an incommensurate propagation wave vector $\textbf{q}_\mathrm{o} = (0,~0,~0.077)_{\mathrm{hex}}$ (Ref.~\onlinecite{16perks2012magneto}). The order manifests itself in X-ray diffraction measurements as a modulation of the Mn2-O bond lengths along the local $x$- and $y$-directions. The negligible variation of the Mn2 valence \cite{16perks2012magneto} renders the orbital order fundamentally different from those in overdoped charge-orbital-ordered perovskite manganites:\cite{LarochellePRB2005,13beale2005orbital,14beale2010xray,15ulbrich2011evidence} Even though preceded by the Mn2-Mn3 charge order (at $T_\mathrm{s}$), the orbital order is incommensurate with all pre-existing periodicity.

At lower temperatures, $\mathrm{CaMn_7O_{12}}$ undergoes two magnetic phase transitions \cite{19zhang2011multiferroic,20johnson2012giant,25Przenioslo1999magnetic} at $T_\mathrm{N1}=90$~K and $T_\mathrm{N2}=48$~K. Since our current study mainly concerns the orbital order, here we refrain from a detailed description of the magnetism in $\mathrm{CaMn_7O_{12}}$, but only mention that (1) the magnetic wave vector $\textbf{q}_\mathrm{m}$ in the $T_\mathrm{N2} < T < T_\mathrm{N1}$ phase locks into the orbital-ordering wave vector, $\textbf{q}_\mathrm{m}=\textbf{q}_\mathrm{o}/2$ (Ref.~\onlinecite{26Slawinski2010structural}), and (2) the orbital order appears to be responsible for the magnetic interactions that lead to the helical magnetic order below $T_\mathrm{N1}$ (Ref.~\onlinecite{16perks2012magneto}). Details about the magnetic phases can be found in Refs.~\onlinecite{16perks2012magneto,20johnson2012giant,25Przenioslo1999magnetic,26Slawinski2010structural,27Przenioslo2013relative} and references therein.

\section{\label{sec:exp}Experimental methods}

Cube-shaped single crystals of $\mathrm{CaMn_7O_{12}}$ were grown by a flux method.\cite{20johnson2012giant} The edges of the cubes (linear dimension about 200 micros) are found to be along the pseudo-cubic $\langle100\rangle$ direction. The crystals have been characterized by X-ray diffraction, resistivity, and specific-heat measurements, all of which indicate that our samples are of high quality. The specific-heat and resistivity measurements were performed using a Quantum Design PPMS. The X-ray diffraction measurement was performed on a Rigaku MiniFlex diffractometer.

Our Raman scattering measurements were performed in a back-scattering geometry using the 632.8~nm line of a He-Ne laser for excitation. The spectra were analyzed using a Horiba Jobin Yvon LabRAM HR Evolution spectrometer, equipped with 600 gr/mm gratings, a liquid-nitrogen-cooled CCD detector, and BragGrate notch filters that allow for measurements down to low wave numbers. The temperature of sample was controlled by a liquid-helium flow cryostat, with the sample under better than 5$\times$10$^{-7}$ Torr vacuum at all times. Consistent Raman spectra have been obtained on several different samples.

\section{\label{sec:exp}Results}
\subsection{\label{sec:char}Characterization measurements}

\begin{figure}[!h]
\includegraphics[width=3in]{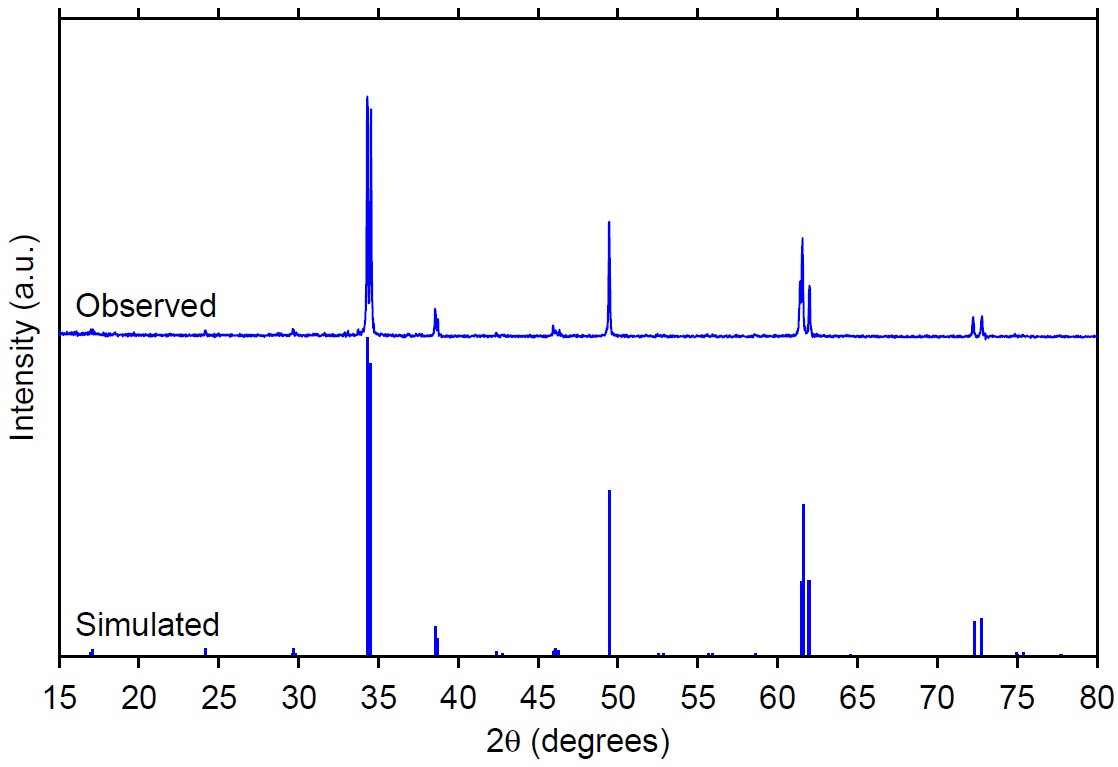}
\caption{\label{fig:XRD}
Powder X-ray diffraction data of $\mathrm{CaMn_7O_{12}}$ obtained at room temperature (measurement and simulation).
}
\end{figure}

Figure~\ref{fig:XRD} displays X-ray diffraction data obtained from a powder sample that was prepared by grinding single crystals. No line pertaining to any impurity phase is found. The lattice parameters in the hexagonal basis are $a = b = 10.4577$~$\mathrm{\AA}$ and $c = 6.3422$~$\mathrm{\AA}$, which are in close agreement with reported values. \cite{17Przenioslo2004charge,19zhang2011multiferroic,23Bochu1980bond}

\begin{figure}[!h]
\includegraphics[width=3in]{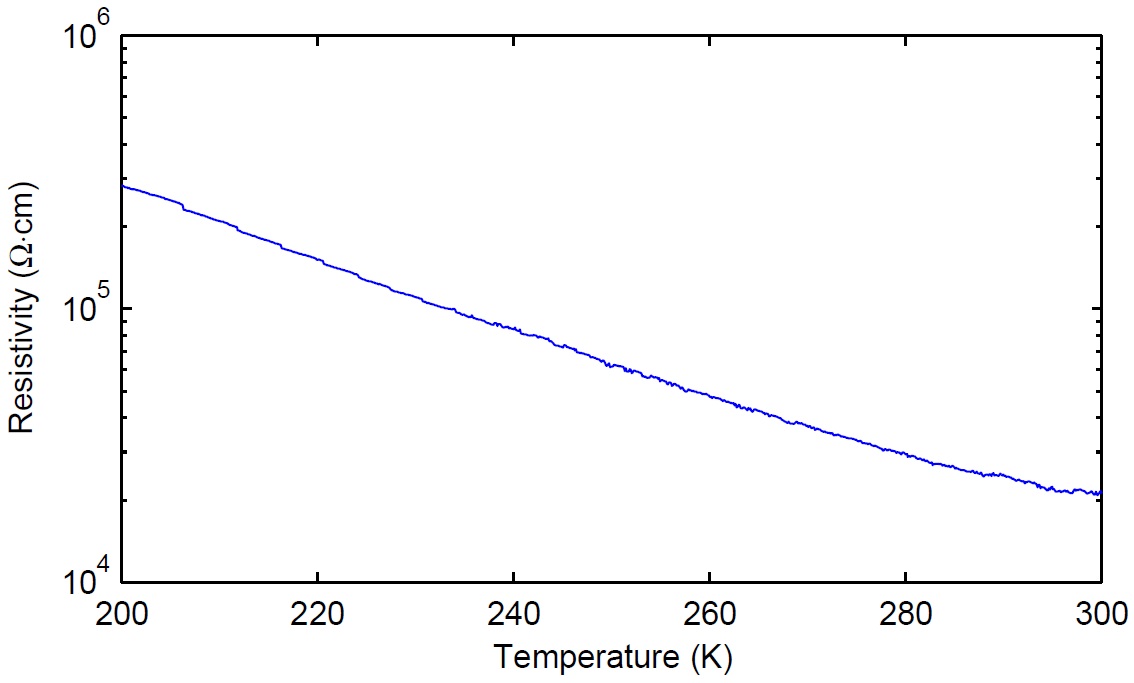}
\caption{\label{fig:RES}
DC resistivity of $\mathrm{CaMn_7O_{12}}$ measured by a two-probe method on a single crystal.
}
\end{figure}

Because our single crystals are small, resistivity measurements were performed by a two-contact method. Electrodes were made on two opposite natural crystal facets (parallel to the $\{100\}$ crystallographic planes of the pseudo-cubic structure) by silver paste. Resistivity values between 200~K and 300~K are displayed in Fig.~\ref{fig:RES}, which are qualitatively consistent, although somewhat larger, than the reported result in Ref.~\onlinecite{36Zeng1999giant} obtained on polycrystalline samples. There is no noticeable anomaly at $T_\mathrm{o}=250$~K. Due to the highly-insulating nature of $\mathrm{CaMn_7O_{12}}$, measurements at lower temperatures are not reliable as the sample resistance approaches the limit of our apparatus.

\begin{figure}[!h]
\includegraphics[width=3in]{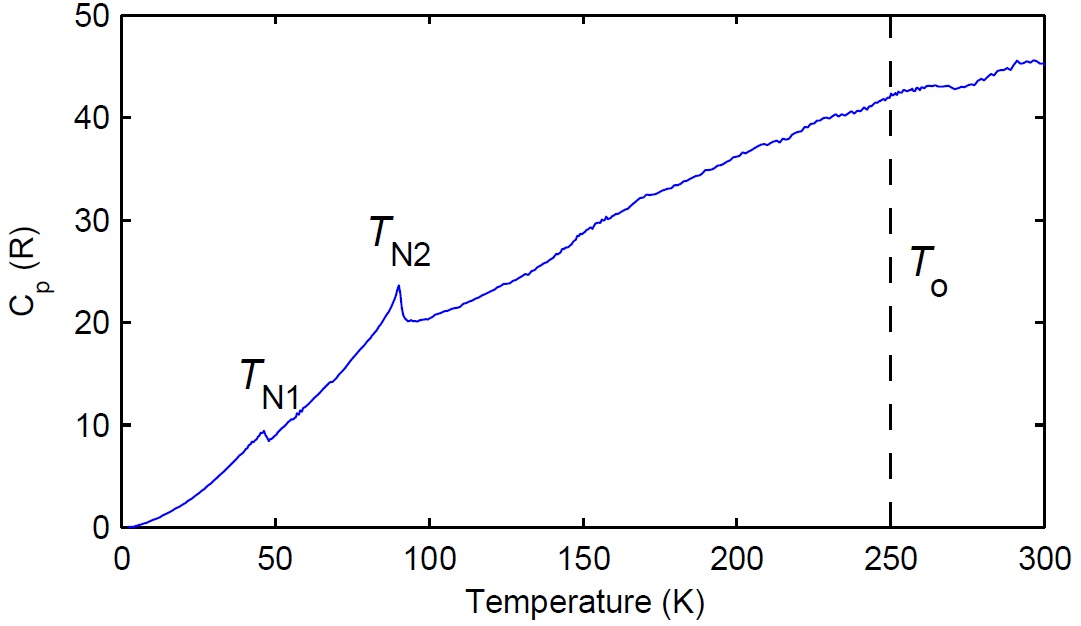}
\caption{\label{fig:HC}
Specific heat of $\mathrm{CaMn_7O_{12}}$. Small variations above 150~K are measurement artifacts related to the thermal properties of the sample mounting we used.
}
\end{figure}

Figure~\ref{fig:HC} displays specific heat of $\mathrm{CaMn_7O_{12}}$ between 4~K and 300~K. A total of 2.0~mg of single crystals were used for the measurement. Clear anomalies can be observed at the $T_\mathrm{N1}$ = 90 K and $T_\mathrm{N2}=48$ K magnetic phase transition temperatures, consistent with an earlier report on polycrystalline samples. \cite{19zhang2011multiferroic} In contrast, no anomaly can be resolved near $T_\mathrm{o}=250$~K.

\subsection{\label{sec:char}Raman spectra of optical phonons}

\begin{figure}[!h]
\includegraphics[width=3in]{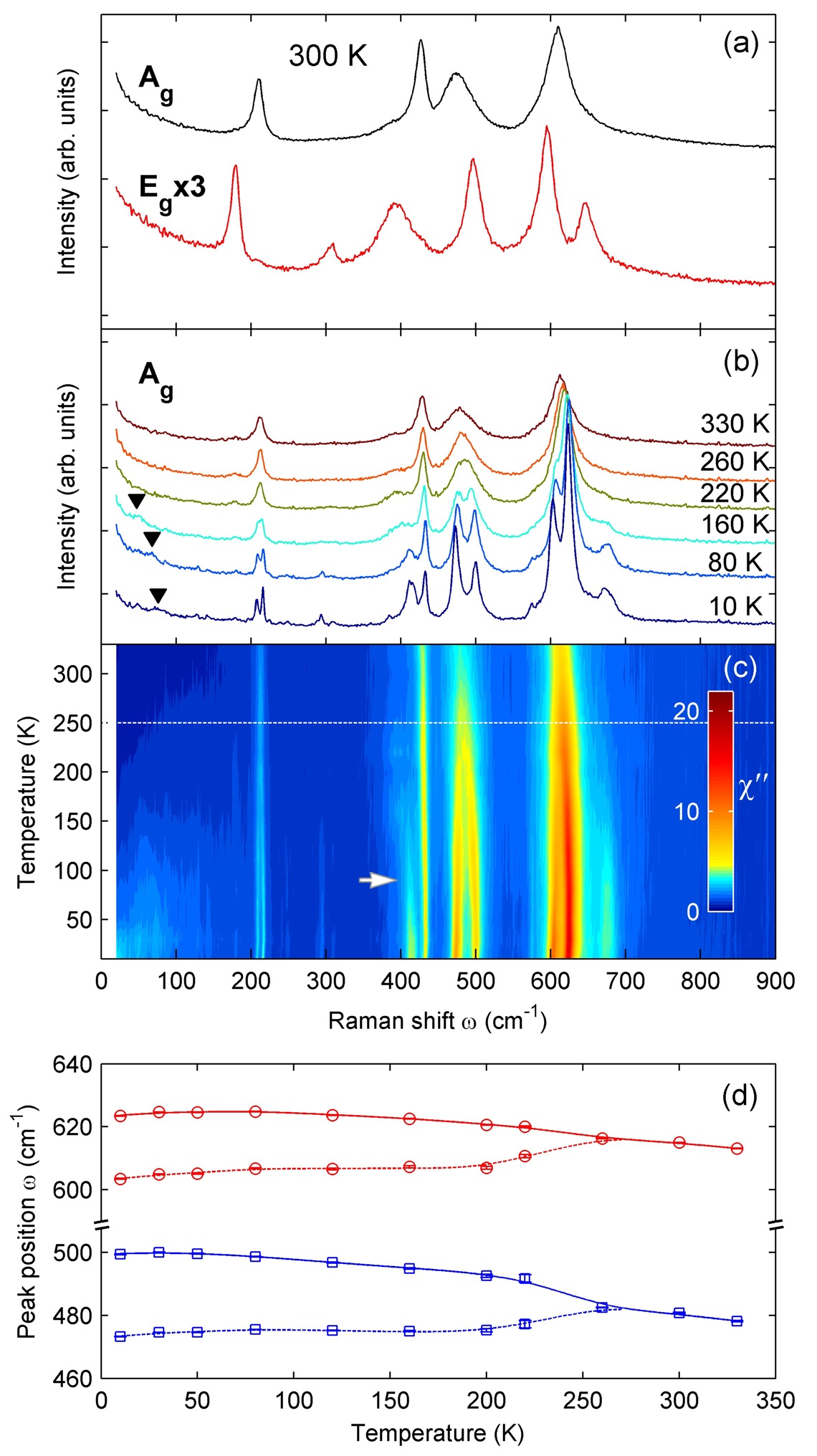}
\caption{\label{fig:OP}
(a) Polarized Raman spectra obtained at 300 K, offset for clarity. Peaks are located at 209.4, 427.3, 473.7, and 610.2 cm$^{-1}$ in the $A_g$ spectrum, and at 180.3, 309.8, 397.8, 496.9, 594.6, and 647.2 cm$^{-1}$ in the $E_g$ spectrum. (b) $A_g$ spectra at selected temperatures, offset for clarity. (c) Color representation of the $A_g$ Raman susceptibility. Dashed line indicates $T_\mathrm{o}= 250$~K. White arrow indicates $T_\mathrm{N1} = 90$~K, below which the peak at 411.3 cm$^{-1}$ gains extra intensity. (d) Positions of prominent phonon peaks near 480 and 610 cm$^{-1}$ as functions of temperature.
}
\end{figure}

We first focus on Raman scattering data that demonstrate the effect of the structural distortions below $T_\mathrm{o}$ on the optical phonon spectrum. According to factor-group analysis, \cite{28rousseau1981normal} there are a total of twelve Raman-active optical phonon modes in the rhombohedral phase of $\mathrm{CaMn_7O_{12}}$: $\Gamma_\mathrm{Raman}$ = 6 $A_g$ + 6 $E_g$, all of which involve oxygen vibrations only. The $A_g$ and $E_g$ modes can be accessed separately using parallel and perpendicular combinations of photon polarizations, respectively, if at least one of the incident and scattered photon polarizations is parallel to the hexagonal $c$-axis. In order to achieve this in our back-scattering experimental geometry, we performed our measurements on polished surfaces parallel to one of the pseudo-cubic $\{110\}$ planes. With a proper choice of this plane, the normal incident laser beam is perpendicular to the hexagonal $c$-axis (the crystals are naturally untwined), which allows for the aforementioned alignment of the photon polarizations. Figure~\ref{fig:OP}a displays polarized Raman data obtained at room temperature. With a negligible cross-leakage between the $A_g$ and $E_g$ signals, the data are in good agreement with our expectation, displaying a total of six (four) prominent peaks in the $E_g$ ($A_g$) geometry. The fact that the total number of distinguishable peaks is less than twelve can be due to the presence of weak and/or nearly degenerate modes. Spectra obtained on natural crystal facets (not shown) exhibit slightly larger cross-leakage between the $A_g$ and $E_g$ signals but somewhat sharper peaks.

\begin{figure}[!h]
\includegraphics[width=3in]{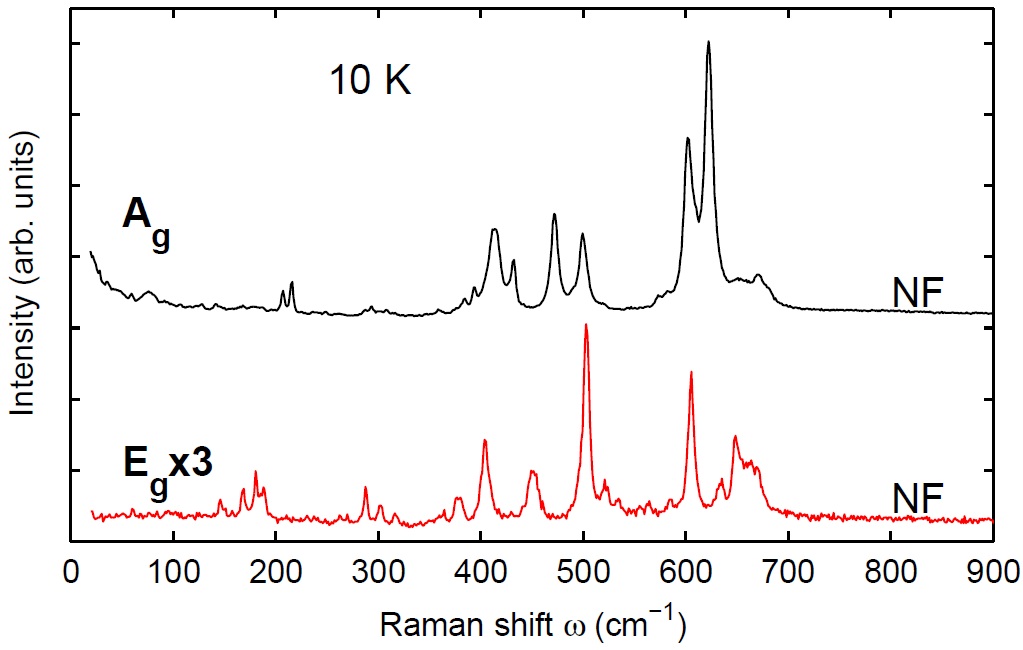}
\caption{\label{fig:AgEg}
Polarized Raman spectra obtained at 10 K. The measurements are performed on natural facets (NF) parallel to the pseudo-cubic $\{110\}$ planes. The $A_g$ and $E_g$ spectra are offset for clarity. The soft mode at $\sim80$~cm$^{-1}$ is absent from the $E_g$ spectrum.
}
\end{figure}

In Fig.~\ref{fig:OP}b-c we present the temperature dependence of the $A_g$ spectrum. Additional peaks appear near all four main peaks upon cooling below $T_\mathrm{o}$. Taking the two high-energy peaks at 300~K as examples, the peak positions near 480 and 610 $\mathrm{cm^{-1}}$, determined by fitting the spectra to Lorentzian profiles, are displayed in Fig.~\ref{fig:OP}d. Since Raman scattering only detects phonons at the Brillouin zone center ($\Gamma$), and because $A_g$ is a one-dimensional representation of the high-temperature structural symmetry group, the systematic appearance of extra peaks below $T_\mathrm{o}$ is most likely due to back-folding of phonon dispersions that gives rise to new optical modes at the $\Gamma$ point. This result is consistent with X-ray diffraction observations of incommensurate structural distortion \cite{16perks2012magneto,17Przenioslo2004charge,18Slawinski2009modulation} that accompanies the orbital order. A closer inspection of the highly accurate data at 10~K (Figs.~\ref{fig:OP}b and \ref{fig:AgEg}) allows us to see many additional features distributed over a wide energy range, which imply further point-group symmetry breaking. As some of the features seem to pick up extra intensities near and below $T_\mathrm{N1}$ (see, $e.g.$, the feature indicated by the white arrow in Fig.~\ref{fig:OP}c), they may be related to inversion-symmetry breaking, which is required for appearance of ferroelectricity below $T_\mathrm{N1}$, and which would mix infrared- and Raman-active modes, although a magnetic origin cannot be completely ruled out at this time.

After completing our measurements, we become aware of a recent Raman study of phonons\cite{29iliev2014raman} in $\mathrm{CaMn_7O_{12}}$. Our data in Figs.~\ref{fig:OP}-\ref{fig:AgEg} are consistent with this report whenever direct comparison can be made. The placement of photon polarization along the hexagonal $c$-axis allows for a much cleaner separation of the $A_g$ and $E_g$ signals in our data compared to the previous report. The authors of Ref.~\onlinecite{29iliev2014raman} did not perform measurements below 100~cm$^{-1}$, which is the main focus in the following.

\subsection{\label{sec:soft}Soft vibrational mode}

\begin{figure}[!h]
\includegraphics[width=3in]{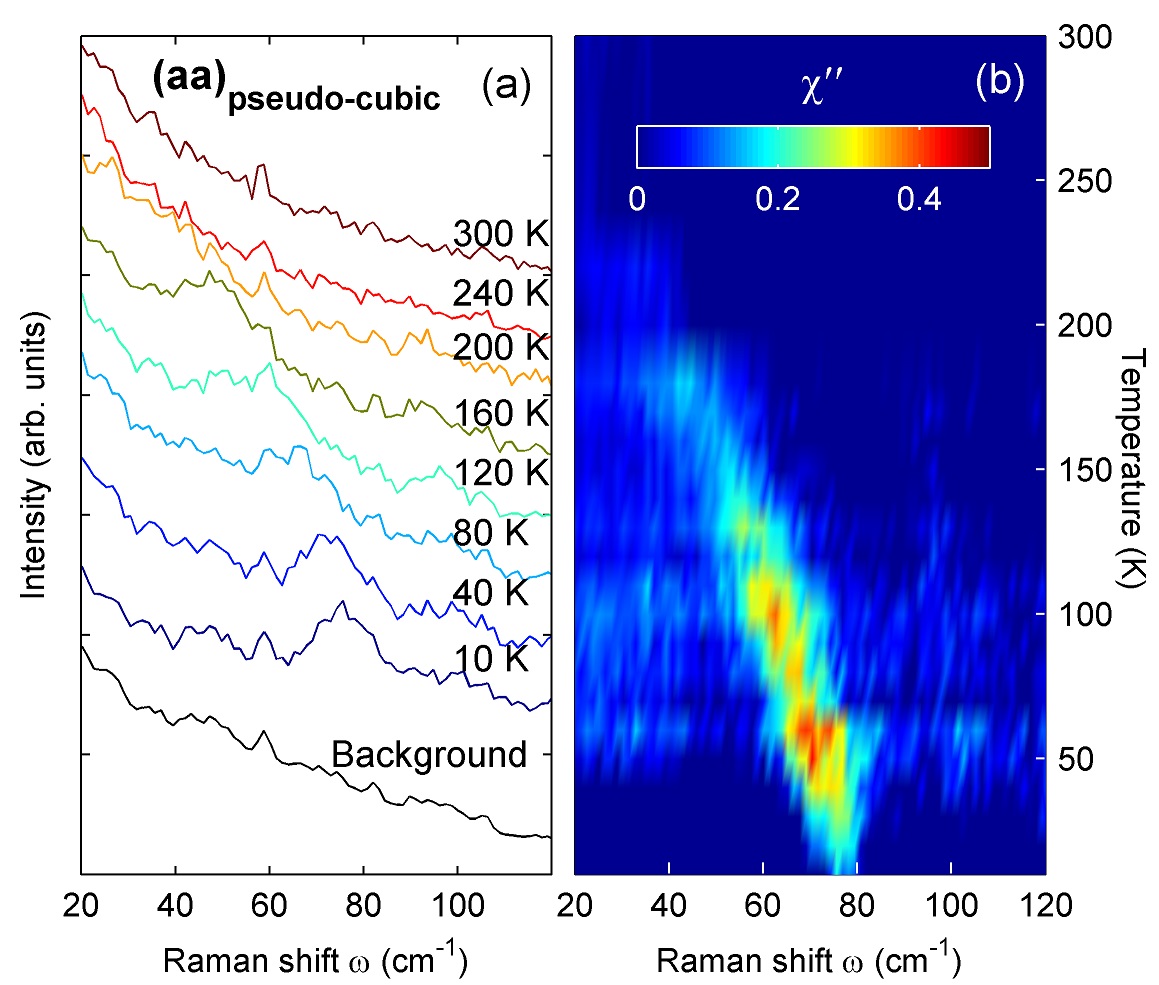}
\caption{\label{fig:SM}
(a) Low-energy Raman spectra at selected temperatures between 10~K and 300~K, offset for clarity. A temperature-independent background contribution due to the optical set up is inferred from the full data set (see text) and displayed at the bottom. (b) Color representation of the temperature dependent the Raman susceptibility.
}
\end{figure}

A close inspection of the data in Fig.~\ref{fig:OP}b allows us to see a low-lying peak (indicated by inverse triangles) which is present only at low temperatures. Since the energy of the peak decreases with increasing temperature, we refer to this feature as the soft mode. Its observation at temperatures above 160~K is difficult from the data in Fig.~\ref{fig:OP}b, because the Rayleigh scattering background is too strong for a clear observation of any features below 50~cm$^{-1}$ (Fig.~\ref{fig:OP}a-b). To overcome this problem, knowing that the soft mode belongs to the $A_g$ representation (it is not observed in the $E_g$ geometry at 10~K, Fig.~\ref{fig:AgEg}), we performed a dedicated set of measurements on large and mirror-like crystal facets parallel to the pseudo-cubic $\{100\}$ planes, focusing on the low-energy part of the spectra. In these measurements the incident and scattered photon polarizations are placed parallel to each other and along one of the pseudo-cubic $\langle100\rangle$ directions, which allows us to probe excitations in both $A_g$ and $E_g$ representations. Raw spectra at selected temperatures are displayed in Fig.~\ref{fig:SM}. With the much suppressed Rayleigh-scattering background due to the improved surface quality, the soft mode is found to emerge below $T_\mathrm{o}$ from the lowest energy of our measurements, and it gradually moves up in energy as temperature is decreased, consistent with the data in Fig.~\ref{fig:OP}.

The observed Raman intensity in Fig.~\ref{fig:SM}a is equal to the imaginary part of the Raman susceptibility ($\chi^{\prime\prime}$) multiplied by the Bose factor, plus a background intensity:
\begin{equation}\label{eqn:one}
I(\omega,T)=\left\{ \frac{1}{1-e^{-\frac{\hbar\omega}{k_\mathrm{B}T}}} \right\}\chi^{\prime\prime}(\omega,T)+B(\omega)
\end{equation}
Here we assume that the background intensity is temperature-independent but energy-dependent, since it is clearly not flat and shows some systematics in the data in Fig.~\ref{fig:SM}a. To determine its value at any given energy $\omega_0$, we fit the temperature-depend intensity $I(\omega_0,T)$ to Eq.~\ref{eqn:one}, assuming that $\chi^{\prime\prime}(\omega_0)$ is independent $T$. The obtained values of $B(\omega_0)$ are plotted as the background curve in Fig.~\ref{fig:SM}a. Even though the soft-mode energy does depend on temperature, and therefore the above assumption about $\chi^{\prime\prime}(\omega_0)$ is only an approximation, we find the estimated background intensities to be in reasonable agreement with all our data. A color representation of the Raman susceptibility $\chi^{\prime\prime}$, after background subtraction and Bose-factor correction, is displayed in Fig.~\ref{fig:SM}b. A softening of the mode towards zero energy is clearly observed as the temperature approaches $T_\mathrm{o}$.

\begin{figure}[!h]
\includegraphics[width=2.8in]{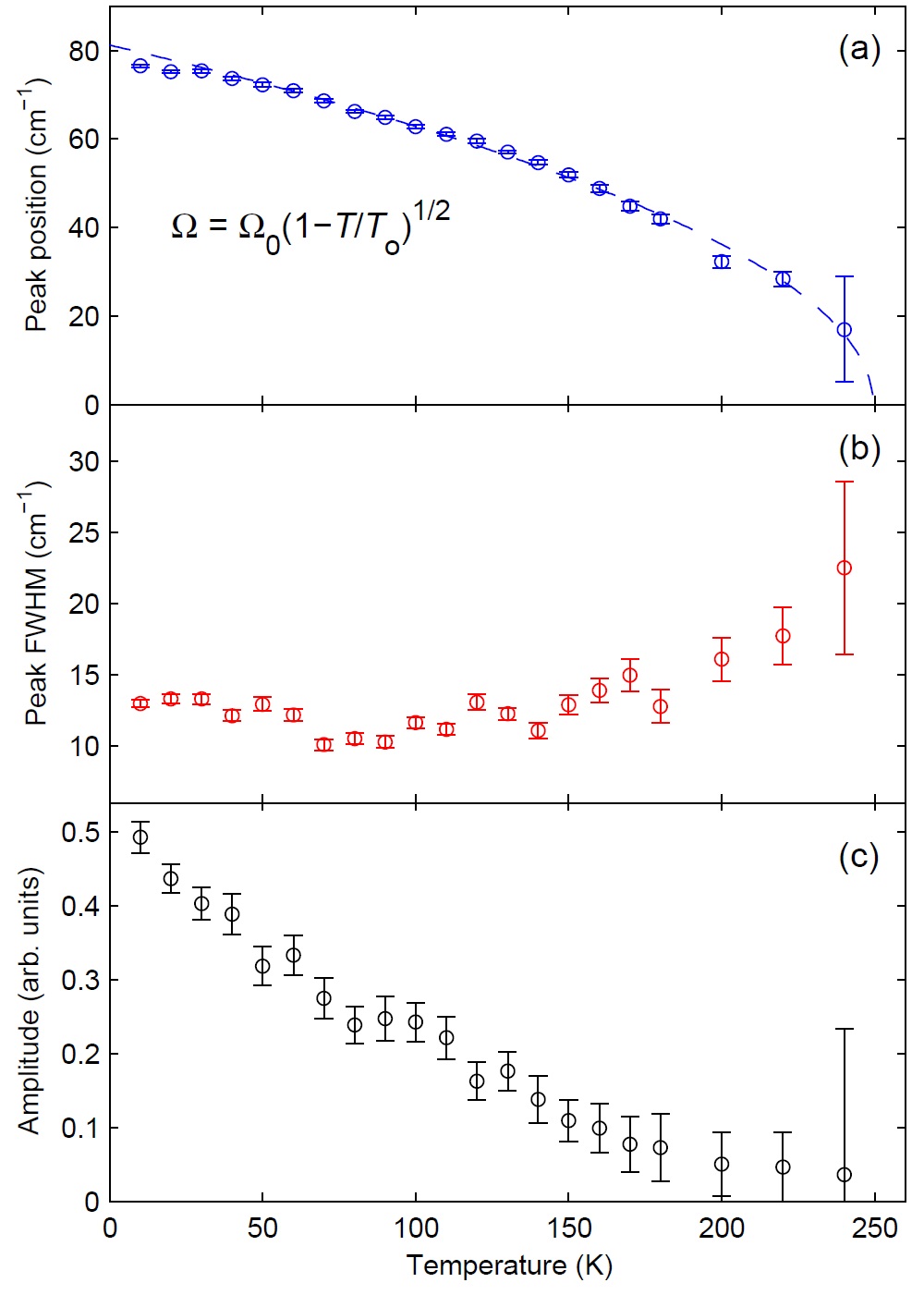}
\caption{\label{fig:fit}
Temperature dependence of the soft-mode energy (a), energy width (b), and Raman-susceptibility amplitude (c). Dashed line in (a) is a power-law fit to the data, resulting in $\Omega_0 = 81.2 \pm 1.2$~cm$^{-1}$ and $T_\mathrm{o} = 249.4 \pm 2.1$~K.
}
\end{figure}

The energy, full width at half maximum (FWHM), and intensity of the soft mode are determined by fitting the signal to a Gaussian profile (Fig.~\ref{fig:fit}). According to Landau's theory for phase transitions, \cite{30poulet1981light} the order-parameter amplitude (see discussion below) is expected to fluctuate at a characteristic frequency proportional to $(1 - T/T_\mathrm{o})^{1/2}$. The data in Fig.~\ref{fig:fit}a can indeed be fitted by $\Omega = \Omega_{0}(1 - T/T_\mathrm{o})^{1/2}$ with $\Omega_0 = 81.2 \pm 1.2$~cm$^{-1}$ and $T_\mathrm{o} = 249.4 \pm 2.1$~K. As the mode softens towards $T_\mathrm{o}$, its energy width increases and the intensity decreases, indicating that it no longer exists just above $T_\mathrm{o}$.

\section{\label{sec:dis}Discussion}

We begin our discussion by noting that the observed soft mode closely resembles related phenomena found in materials that exhibit structural phase transitions driven by soft phonons. \cite{31scott1974soft,32cummins1990experimental,33Gruner1994Density} In such materials, the frequency (or energy) of a phonon decreases as the transition temperature $T_\mathrm{c}$ is approached from above, and it reaches zero at $T_\mathrm{c}$, where the corresponding movement of the atoms freezes in. The resultant lattice distortion breaks at least one of the structural symmetries of the high-temperature phase, and the distortion amplitude continuously increases from zero upon further cooling below $T_\mathrm{c}$. The amplitude increase can be phenomenologically described by Landau's theory: $A = A_{0}(1 - T/T_\mathrm{c})^{1/2}$, where $A_{0}$ is the fully-saturated amplitude of the distortion. Since variation of $A$ from its mean value costs energy, the variations in the long-wavelength limit exhibit a characteristic frequency, which increases with decreasing temperature in the same fashion as $A$ itself.\cite{30poulet1981light} This frequency defines a distinct type of excitations involving the amplitude of the order parameter, which is commonly referred to as the amplitude excitations, or ``amplitudons''.\cite{31scott1974soft,32cummins1990experimental,33Gruner1994Density} They are always Raman-active below $T_\mathrm{c}$ and belong to the $A_g$ totally symmetric representation. \cite{32cummins1990experimental} Since all of these attributes are consistent with what we find, we conclude that the soft mode in $\mathrm{CaMn_7O_{12}}$ is the amplitude excitation of the order parameter, which has a combined orbital and lattice character.\cite{16perks2012magneto} To our knowledge, this is the first observation of amplitude excitations associated with an orbital order.

Our observation of the amplitude excitation unambiguously proves that the phase transition at $T_\mathrm{o}$ is of second-order nature, which has remained hitherto undetermined. Moreover, the emergence of the soft mode from zero energy at $T_\mathrm{o}$ suggests that the orbital order is triggered by a soft-phonon-driven lattice instability at $\mathbf{q_\mathrm{o}}$ in reciprocal space, as will be further discussed below. We are able to track the soft mode up to temperatures very close to $T_\mathrm{o}$ (Fig.~\ref{fig:fit}a), where the softening is as large as 70\%. This large softening suggests that $T_\mathrm{o}$ is very close to the mean-field transition temperature, presumably due to the three-dimensional nature of $\mathrm{CaMn_7O_{12}}$.

While the incommensurate orbital order has been suggested to be a key to the realization of the magnetic phases at lower temperatures, \cite{16perks2012magneto} here we do not observe any change in the characteristics of the soft mode across $T_\mathrm{N1}= 90$~K or $T_\mathrm{N2} = 48$~K. This is not an entirely unexpected result, since neither the orbital-ordering wave vector \cite{18Slawinski2009modulation} nor the intensity of the satellite diffraction peaks \cite{17Przenioslo2004charge} are found to exhibit any anomaly at these temperatures. The lack of such anomalies indicates that the orbital order is well-developed and robust at the magnetic transition temperatures, and that the feedback from the magnetic ordering is not strong enough to cause a noticeable effect on the orbital order.

When a structural phase transition is driven by the softening of a zone-center phonon, it is possible to detect the phonon with Raman scattering already above the transition temperature. \cite{31scott1974soft} In $\mathrm{CaMn_7O_{12}}$, however, because the orbital order occurs at a finite momentum $\mathbf{q}_\mathrm{o}$, the associated phonon is not Raman-active above $T_\mathrm{o}$. Detecting the related phonon anomalies and determining the associated Eigenvectors above $T_\mathrm{o}$ will provide useful information on the microscopic mechanism of the orbital order, but will require momentum-resolved spectroscopic techniques such as X-ray and neutron scattering, the latter of which requires large single crystals. Our present study nevertheless provides useful information for such future investigations. Assuming that the low-temperature frequency $\Omega_0$ of the soft mode is related to that of a phonon in the normal state at $\mathbf{q_\mathrm{o}} = (0, 0, 0.077)_{\mathrm{hex}}$ up to a factor on the order of unity, \textit{i.e.}, similar to the relationship between amplitude excitations and phonon anomalies in the materials reviewed in Refs.~\onlinecite{31scott1974soft,32cummins1990experimental,33Gruner1994Density}, we can infer that the anomalous phonon has to belong to an optical branch, because $\Omega_0$ is unrealistically high for an acoustic mode not far from the $\Gamma$ point. Moreover, our specific-heat measurement detects no clear anomaly at $T_\mathrm{o}$ (Fig.~\ref{fig:HC}), which implies that the phonon anomalies are confined to a small region in reciprocal space and/or exist only in low-lying phonon branches that are already well-populated at $T_\mathrm{o}$, so that the softening near $T_\mathrm{o}$ does not substantially affect the total specific heat.

Several microscopic mechanisms have been suggested for incommensurate structural phase transitions. They mainly concern two gross categories of materials. The first category consists of insulators, for which the most invoked mechanisms commonly involve some type of competition, \textit{e.g.}, between long- and short-range lattice interactions. \cite{32cummins1990experimental} The second category consists of metallic charge-density-wave (CDW) materials, \cite{33Gruner1994Density} in which the phase transition is driven by Fermi-surface (FS) nesting in conjunction with electron-phonon coupling; the CDW wave vector and phonon (Kohn) anomalies \cite{34pouget1991neutron,35hoesch2009giant} are closely related to the nesting wave vector $2\mathbf{k_\mathrm{F}}$. Since $\mathrm{CaMn_7O_{12}}$ is highly insulating \cite{36Zeng1999giant} both above and below $T_\mathrm{o}$ with no resistivity anomaly at $T_\mathrm{o}$ (Fig.~\ref{fig:RES}), it appears at first sight that one should seek for a mechanism similar to those for the insulators. As for the competing interactions, corporative Jahn-Teller effects might lead to geometric frustration among distortions on neighboring MnO$_6$ octahedra and MnO$_4$ rhombuses, but the exact form of interactions and frustration pathway remain unclear at this point given the rather complicated network of Mn ions (Fig.~\ref{fig:structure}a).

From a different perspective, the incommensurate lattice instability in $\mathrm{CaMn_7O_{12}}$ may be caused by prominent $\mathbf{q}$-dependent electron-phonon coupling. This picture differs from the above in that its starting point is in reciprocal rather than real space, and it bears some similarity to the CDW mechanism, albeit the selection of $\textbf{q}_\mathrm{o}$ is not due to FS nesting. A strong motivation for such a starting point is that phonon anomalies in $\mathrm{CaMn_7O_{12}}$ are likely localized in $\mathbf{q}$ as discussed above, which is in stark contrast to pronounced specific-heat anomalies and phonon softening over broad momentum regions in ionic insulators. \cite{32cummins1990experimental,37LopezEcharri1980specific,38iizumi1977structural} Sharp anomalies in the reciprocal space can only be produced by long-range interactions in real space, hence a momentum-based starting point will likely prove more convenient and physical.

It is well-known that phonons can be coupled not only to the electron density, but also to the orbital state of the electrons. \cite{39millis1996fermi} The latter can be realized, in particular, via the Jahn-Teller effect, which is evidently strong in the manganites. In fact, even for the quasi-two-dimensional CDW material NbSe$_2$, in which the FS nesting vector clearly differs from the CDW wave vector, \cite{40Borisenko2009two,41johannes2006fermi} it has been suggested that $\mathbf{q}$-dependent electron-phonon coupling is the determining factor for the wave vector of the CDW instability. \cite{41johannes2006fermi,42weber2011extended} While the electron-phonon coupling in NbSe$_2$ is primarily in the electron-density sector, $\mathrm{CaMn_7O_{12}}$ may be the first example of electron-phonon coupling in the orbital sector being the main driving force towards an orbital-ordered state.

Last but not least, we speculate on a related topic based on our result. Recently, incommensurate charge ordering phenomena have been observed in high-temperature cuprate superconductors. \cite{43wu2011magnetic,44Ghiringhelli2012long} Evidence for periodic variations in the occupation of the planar oxygen $p_x$ and $p_y$ orbitals is found, \cite{45comin2014the,46Fujita2014direct} along with seemingly incompatible observations of weak FS nesting \cite{47comin2014charge} and unusually sharp phonon anomalies. \cite{48letacon2014inelastic} Since the $3d^9$ electronic configuration of Cu$^{2+}$, similar to the $3d^4$ configuration of Mn$^{3+}$, is prone to strong electron-phonon coupling via the Jahn-Teller effect, a mechanism similar to the one outlined above may be relevant to the high-$T_\mathrm{c}$ cuprates as well.

\section{\label{sec:con}Conclusions}

In summary, we report a systematic study of lattice vibrations in $\mathrm{CaMn_7O_{12}}$ using Raman spectroscopy. We observe additional optical phonon peaks in the orbital-ordered state, which we attribute to Brillouin-zone folding as a result of the incommensurate structural distortion that accompanies the orbital order. A new soft vibrational mode appears below the orbital-ordering temperature, with a canonical power-law temperature-dependent energy that can be very well described by Landau's theory. This mode can be regarded as the amplitude excitation of the composite order parameter that involves both orbital and lattice degrees of freedom, and its observation demonstrates that the transition is of second-order nature. The fact that the soft mode emerges from zero energy at $T_\mathrm{o}$ strongly suggests that the orbital order is triggered by a soft-mode-driven lattice instability.

The softening of phonon modes above $T_\mathrm{o}$ can be explained either by competing interactions in real space or by $\mathbf{q}$-dependent electron-phonon coupling in reciprocal space. While Raman scattering does not detect the softening above $T_\mathrm{o}$, our complimentary measurement of the specific heat suggests that the softening is confined to restricted regions of $\mathbf{q}$, which is in favor of the latter explanation. Further verifications of this using momentum-resolved spectroscopies is desirable for future studies.

\begin{center}
$\textbf{Acknowledgments}$
\end{center}

We wish to thank Mathieu Le Tacon, Fa Wang, Ji Feng, Martin Greven, Shuai Dong, Junming Liu, Hongjun Xiang, and Naoto Nagaosa for stimulating discussions, and Yan Zhang, Chenglong Zhang, and Shuang Jia for assistance on characterization measurements. This work is supported by the NSF of China (No. 11374024) and the NBRP of China (No. 2013CB921903). R.Y. and L.D. are supported in part by the President's Fund for Undergraduate Research of Peking University.

\bibliography{CMO_arXiv}

\end{document}